\documentclass[10pt,aps,prb,twocolumn,showpacs]{revtex4-1}
\usepackage{amsmath}
\usepackage{setspace}
\usepackage{dcolumn}
\usepackage{natbib}
\usepackage[version=3]{mhchem}
\usepackage[utf8]{inputenc}

\begin{document}

\title{Comment on ``First-principles study of the influence of (110)-oriented
strain on the ferroelectric properties of rutile \ce{TiO2}''}
\author{Keith Refson}
\email{Keith.Refson@stfc.ac.uk}
\affiliation{STFC, Rutherford Appleton Laboratory, Didcot, Oxfordshire, OX11 0QX, UK}
\author{Barbara Montanari}
\affiliation{STFC, Rutherford Appleton Laboratory, Didcot, Oxfordshire, OX11 0QX, UK}
\author{Pavlin D. Mitev}
\author{Kersti Hermansson}
\affiliation{Materials Chemistry, {\AA}ngstr\"{o}m Laboratory, P.O. Box 538, SE-752 21 Uppsala, Sweden}
\author{Nicholas M. Harrison}
\affiliation{Department of Chemistry,Imperial College London, South Kensington Campus, London, SW7 2AZ, UK}
\altaffiliation{STFC Daresbury Laboratory, Cheshire, WA4 4AD, UK}

\begin{abstract}
  In a recent article, Gr\"{u}nebohm \textit{et al.} [Phys.
  Rev. \textbf{B} 84 132105 (2011)] report that they fail to reproduce
  the A$_{2u}$ ferroelectric instability of \ce{TiO2} in the rutile
  structure calculated with density functional theory within the
  PBE-GGA approximation by Montanari \textit{et al.} [Chem. Phys. Lett
  \textbf{364}, 528 (2002)].  We demonstrate that this disagreement arises 
  from an erroneous treatment of \ce{Ti} $3s$ and $3p$ semi-core electrons as core in 
  their calculations. Fortuitously the effect of the frozen semi-core 
  pseudopotential cancels the phonon instability of the PBE exchange-correlation, 
  and the combination yields phonon frequencies similar to the LDA harmonic values.
  
  Gr\"{u}nebohm \textit{et al.} also attempted and failed to reproduce the soft acoustic 
  phonon mode instability under (110) strain reported by Mitev \textit{et al.} [Phys. Rev. \textbf{B} 81 134303 (2010)].
  For this mode the combination of PBE-GGA and frozen semi-core 
  yields a small imaginary frequency of 9.8i.
  The failure of Gr\"{u}nebohm \textit{et al.} to find this mode probably arose from numerical limitations 
  of the geometry optimization approach in the presence of a shallow double well potential; the optimization 
  method is not suitable for locating such instabilities.

\end{abstract}


\maketitle

\section{Introduction}

Rutile \ce{TiO2} is an incipient ferroelectric material. Over a 20 year
period density functional theory has been used to develop an
understanding of its dielectric response and lattice dynamical
properties. It has also established the exquisite sensitivity of these
properties to strain, volume changes and thus to approximations such
as the treatment of electronic exchange and correlation\cite{Montanari2004}.

In 2002 two of us demonstrated\cite{Montanari2002} that within the
Perdew Burke and Enzerhof generalized gradient approximation
(PBE-GGA)\cite{perdew1997} the A$_{2u}$ mode is unstable at the
equilibrium geometry. The resulting structural instability
corresponds to a ferroelectric distortion of the crystal in
qualitative disagreement with the observed properties of rutile
\ce{TiO2}\footnotemark[2].
\footnotetext[2]{The shallow double well potential obtained using PBE-GGA
yields a prediction of a mechanical instability in the classical zero-temperature ``athermal'' limit.
However this ignores nuclear zero-point quantum effects and thermal 
motion. It is possible that a fully quantum statistical mechanical PBE calculation which 
included these effects would predict an undistorted rutile structure even at low temperature.}
In the local density approximation (LDA)\cite{Perdew1981}
the equilibrium cell volume is slightly smaller and a stable structure
is obtained, giving a quantitatively correct description of the
lattice dynamics and dielectric properties.  This motivated the choice
of LDA functional for our study of
Ref.~\onlinecite{mitev2010} on the influence of strain on the
ferroelectric properties and phonon modes of rutile \ce{TiO2}. The
major conclusion of that work was the existence of a low frequency
acoustic phonon branch over a broad region of the Brillouin Zone
around (1/2,1/2,1/4) which became soft under negative pressure or
anisotropic strain.

In article Ref.~\onlinecite{Grunebohm2011}, Gr\"{u}nebohm, Ederer,
and Entel, published another study on the same topic, in which they attempted and
failed to reproduce those central findings of our work. 
Specifically they report that that the PBE-GGA does not predict a phonon
instability of the rutile structure, that there is no softening of the
A$_{2u}$ mode under (110) strain, nor is there a destabilization under
(110) strain at (1/2,1/2,1/4) by a soft acoustic mode.

It would be highly unsatisfactory for this discrepancy to remain
unresolved in the literature; it would weaken the basis
for future studies on this topic, and undermine
confidence in the reproducibility of DFT plane-wave methods for this type
of study.  Gr\"{u}nebohm \textit{et al.} offer a speculation that the discrepancy
arises from ``the different potentials used and the resulting
difference in lattice parameters''. We tested that hypothesis, and
confirm it as the most likely origin of the discrepancy.
We show that the error lies in the choice of pseudopotentials used by Gr\"{u}nebohm \textit{et
al.} specifically their treatment of the \ce{Ti} $3s$ and $3p$ semicore electrons as part
of the frozen pseudopotential core instead of as valence.

Textbook wisdom holds that in the case of titanate perovskites the
frozen semicore approximation introduces significant errors in
ferroelectric properties and that $3p$ and possibly $3s$ states must be
explicitly included\cite{*[{p.~265 in }] [{}] martin2004}.  Indeed
most pseudopotential studies of \ce{TiO2} and titanate perovskites
published since the mid 1990s have treated the \ce{Ti} $3s$ and $3p$ electrons
as valence, claiming that the frozen semicore approximation is
inaccurate. However there is no report containing quantitative evidence of the
consequences for phonon frequencies.

\begin{table}

\caption{\label{table:pseudopotentials}Pseudopotential configurations used.  A single ultrasoft
projector was used for each core or semi-core state. Two ultrasoft
projectors were used for each valence state except for UF1 where a
single projector was used for \ce{Ti} 4s and UF3 where a single
norm-conserving projector was used for \ce{Ti} 4s.  For all frozen core
and semi-core states a pseudized core charge\cite{Louie1982} was used.}

\begin{ruledtabular}
\begin{tabular}{p{1cm}p{2.25cm}p{1.4cm}p{1cm}}
Label& UF1  &UF2  &UF3  \\\hline
   \ce{Ti} ref.  &  $2s^{2}2p^{6}3s^{2}3p^{6}4s^{2}3d^{2}$ &  $3s^{2}3p^{6}4s^{2}3d^{2}$ & $4s^{2}3d^{2}$ \\
   $r_{c}$      &  1.2/1.2/1.8             & 1.8              & 2.2 \\
   $r_\text{inner}/a_{0}$  & 0.5                      & 1.5              & 1.3 \\
   $l_\text{loc}/a_{0}$    & $d$                        & $d$                & $s$ \\
   O ref.   & $1s^{2}2s^{2}2p^{4}$              & $2s^{2}2p^{4}$          & $2s^{2}2p^{4}$ \\
   $r_{c}$/$a_{0}$       & 0.8                      & 1.3              &  1.3  \\
   $r_\text{inner}/a_{0}$   & 0.5                      & 0.9              & 0.9 \\
   Cutoff /eV      & 1300                     & 750              & 750
\end{tabular}
\end{ruledtabular}
\end{table}

\begin{table}

\caption{\label{table:structure} Calculated structural parameters of rutile \ce{TiO2}}
\begin{ruledtabular}
\begin{tabular}{lddd}
               &    \multicolumn{1}{c}{a/A}   &     \multicolumn{1}{c}{c/A}   &     \multicolumn{1}{c}{u} \\ \hline
PW-LDA (UF1)   & 4.550 &  2.919 &  0.3039 \\
PW-LDA (UF2)   & 4.549 &  2.919 &  0.3039 \\
PW-LDA (UF3)   & 4.597 &  2.902 &  0.3023 \\
LCAO-LDA\footnotemark[1]   & 4.548 &  2.944 &  0.305 \\
FP-LAPW-LDA\footnotemark[2] & 4.558 &  2.920 &  0.3039 \\\hline
PW-GGA (UF1)   & 4.642 &  2.963 &  0.3051 \\
PW-GGA (UF2)   & 4.642 &  2.964 &  0.3051 \\
PW-GGA (UF3)   & 4.684 &  2.953 &  0.3035 \\
PW-GGA\footnotemark[3]  & 4.641 &  2.966 &  0.305 \\
PW-GGA\footnotemark[4]   & 4.664 &  2.969 &  0.3047 \\
LCAO-GGA\footnotemark[5]   & 4.623 &  2.987 &  0.306
\end{tabular}
\end{ruledtabular}
\footnotetext[1]{Ref.~\onlinecite{Muscat2002,Montanari2004}}
\footnotetext[2]{Ref.~\onlinecite{Cangiani2003}}
\footnotetext[3]{Ref.~\onlinecite{Montanari2002}}
\footnotetext[4]{Ref.~\onlinecite{Grunebohm2011}}
\footnotetext[5]{Ref.~\onlinecite{Muscat1999}}
\end{table}

Ghosez \textit{et al.}\cite{Ghosez1995,Ghosez1998} performed band-by
band decomposition of the Born effective charges and showed that the
Ti $3p$ orbitals contribute a value of -0.22e to the effective charge on
the \ce{Ti} atoms in \ce{SrTiO3} and -0.43e in the case of \ce{BaTiO3}
perovskites.  The same authors also argue\cite{Ghosez1997} that small
contributions to this effective charge are responsible for the
destabilization of the cubic phase by the softening of an optic phonon
and a phase transition to the rhombohedral phase. Tegner \textit{et
al.}\cite{tegner2011} performed a comprehensive investigation on the
effect of inclusion of \ce{Ti} $3s$ and $3p$ on cohesive energies, phase
stability and electronic density of states of metallic Ti, concluding
that frozen semicore \ce{Ti} pseudopotentials result in poor
transferability and that the relative stability of the $\omega$ and
hcp phases is reversed.  Deskins\cite{AaronDeskins2009} found an
0.3~eV difference in the cohesive energy of \ce{TiO2} and again that
semicore \ce{Ti} pseudopotentials exhibit worse transferability in the
case of changes of oxidation state, with errors of up to 0.1 eV in
dissociation reaction energies of \ce{TiCl$_{n}$} and absorption
energetics of a variety of molecules and atoms on rutile \ce{TiO2}
surfaces. Holzwarth \textit{et al.}\cite{Holzwarth1997} highlighted a
large error in the atomic configuration energy for s-d promotion in Ti
atoms of 90~meV obtained within the frozen-semicore approximation.
However Perron \textit{et al.}\cite{Perron2007} showed that the effect
of a frozen-core pseudopotential on \emph{structural} parameters of
\ce{TiO2} rutile and anatase is less than 0.5\% provided that a
nonlinear core correction term is included\footnotemark[1]. 

The question of how many electrons to treat as valence is independent
of other aspects of the pseudopotential construction or formalism, and
holds equally for norm-conserving, ultrasoft or PAW
pseudopotentials\cite{Holzwarth1997}, as it concerns the inclusion or
omission of the physics of polarizability of the \ce{Ti} $3s$ and $3p$
electrons.

In summary, previous investigations are consistent with a picture
in which structural parameters are little affected by the treatment of
$3s$ and $3p$, relative energies of phases and cohesive energies are
noticeably affected. Frozen semicore pseudopotentials have
poor transferability resulting in significant errors in
dielectric properties and chemical reaction energetics.  As no
results have been reported on the consequence for phonon frequencies
and mechanical stability, we present them here.

\section{GGA Instability of rutile structure}

\begin{table*}

\caption{\label{table:prim-frequencies} Calculated and experimental $\Gamma$-point phonon frequencies of bulk rutile \ce{TiO2} (cm$^{-1}$).  (TO frequencies only.) }

\begin{ruledtabular}
\begin{tabular}{ldddddddddd}
Mode & \multicolumn{1}{c}{LDA-UF1} & \multicolumn{1}{c}{LDA-UF2} & 
\multicolumn{1}{c}{LDA\footnotemark[1]} & \multicolumn{1}{c}{LDA-LAPW\footnotemark[2]} &  \multicolumn{1}{c}{PBE-UF1} &   \multicolumn{1}{c}{PBE-UF2}  & \multicolumn{1}{c}{PBE\footnotemark[1]} & \multicolumn{1}{c}{PBE-UF3\footnotemark[3]} & \multicolumn{1}{c}{PBE-UF3} & \multicolumn{1}{c}{Expt\footnotemark[1]} \\ \hline
B$_{2u}$ &      99.2 &      101.7 &     104.0 &      111.5 &    31.0i &     28.4i &     79.2 &    96.1   &     91.2 &    113 \\
A$_{2u}$ &     125.1 &      128.7 &     154.4 &      153.0 &    89.6i &    106.6i &     86.3i&   143.7   &    115.7 &    142 \\
E$_{u}$  &     135.2 &      138.8 &     191.4 &      156.4 &    69.8i &     62.5i &    124.0 &   137.7   &    111.2 &    189 \\
B$_{2g}$ &     136.9 &      137.1 &     137.0 &      138.2 &    127.1 &     150.5 &    154.2 &   131.2   &    144.4 &    142 \\
E$_{u}$  &     381.5 &      382.4 &     383.9 &      384.8 &    353.8 &     355.1 &    353.5 &   383.2   &    366.4 &    388 \\
B$_{2u}$ &     390.9 &      392.4 &     393.0 &      398.2 &    362.3 &     356.4 &    357.5 &   400.5   &    380.3 &    406 \\
A$_{2g}$ &     417.6 &      419.5 &     421.7 &      416.0 &    408.5 &     417.9 &    423.6 &   407.6   &    405.6 &     - \\
E$_{g}$  &     463.1 &      463.5 &     463.2 &      463.7 &    429.1 &     428.9 &    429.2 &   472.2   &    464.3 &    455 \\
E$_{u}$  &     486.7 &      487.3 &     488.4 &      487.0 &    469.6 &     469.7 &    488.4 &   480.0   &    491.4 &    494 \\
A$_{1g}$ &     609.9 &      610.4 &     611.6 &      612.0 &    567.8 &     567.3 &    565.9 &   626.9   &    611.1 &    610 \\
B$_{1g}$ &     817.3 &      818.2 &     824.7 &      816.1 &    770.8 &     771.4 &    774.3 &   815.2   &    794.6 &    825
\end{tabular}
\end{ruledtabular}
\footnotetext[1]{Ref.~\onlinecite{Montanari2002}}
\footnotetext[2]{Ref.~\onlinecite{Cangiani2003}. These are
  pseudopotential-LAPW, not all-electron FP-LAPW calculations}
\footnotetext[3]{Calculated using the optimized lattice parameters of UF2}
\end{table*}

We performed calculations on bulk rutile \ce{TiO2} using the CASTEP
code\cite{clark2005}, with a series of pseudopotentials to test the
effect of the frozen-core approximation on successive shells. The
pseudopotentials were of the ultrasoft variety generated using
Vanderbilt's method\cite{vanderbilt1990}. Small core radii were chosen
to generate accurate but ``hard'' pseudopotentials (\emph{i.e.} which
require a large cutoff) for both elements to leave the frozen-core
approximation as the dominant contribution to the error.  Details of
these potentials are listed in table~\ref{table:pseudopotentials}.
Three sets of pseudopotentials are designated UF1, UF2, and UF3
denoting that the Ti~$1s$, $2s$ and $2p$, and $3s$ and $3p$ shells were frozen
respectively.  The UF3 set corresponds to the ``large core'' set used
by Gr\"{u}nebohm \textit{et al.} and the UF2 set corresponds to the
``small core'' set used by Montanari and Harrison and by previous
authors.  The UF1 set gives a ``nearly all electron'' calculation in
which only the \ce{Ti} $1s$ orbitals are in the core, and \ce{Ti} $2s$ and $2p$ and O
$1s$ are explicitly treated as valence.  The computational cost of
treating core states as valence is surprisingly reasonable because the
sharply-peaked core orbitals are represented mostly by the pseudized
augmentation functions of the Vanderbilt scheme.  A very fine FFT grid
is required only for the electron density to represent the
augmentation charge, not for the soft orbital functions. This ``all
electron pseudopotential'' approach has previously used successfully
for calculations on iron at TPa pressures\cite{ISI:000271090200005}.

In each case a geometry optimization was performed to optimize both
lattice parameters and the internal co-ordinate, followed by a finite
displacement calculation of the dynamical matrix. This was
diagonalized to yield the gamma-point phonon frequencies.  The
Brillouin zone was sampled using a Monkhorst Pack\cite{Monkhorst1976}
grid of dimensions $4 \times 4 \times 6$ and plane-wave cutoffs for
each USP set are listed in table~\ref{table:pseudopotentials}.  These
choices yield a convergence of better than 1~meV/atom in total energy,
0.00025~A in lattice parameter. The geometry was relaxed until the
force residual on each atom was less than 0.005~eV/A.  For most
phonons the frequencies were converged to better than 0.1~cm$^{-1}$; 
for the most sensitive phonon modes (which become imaginary) the maximum error was 3~cm$^{-1}$.

The calculated structural parameters are shown in
table~\ref{table:structure}.  LDA lattice parameters for UF1 and UF2
are within 0.2\% of those calculated using the reference all-electron
techniques\cite{Montanari2004,Cangiani2003}.  Freezing the \ce{Ti} semicore
states slightly increases the lattice parameters but by less than 1\%,
consistent with the values reported by Gr\"{u}nebohm \textit{et al.}
and by previous studies\cite{Perron2007}$^{,}$\footnotemark[1].
\footnotetext[1]{A much larger discrepancy between small- and
  large-core pseudopotentials in early calculations was shown to be
  due to the omission of the ``non-linear core correction''
  term\cite{Louie1982}, essential for the correct treatment of
  exchange and correlation in the presence of core-valence overlap in
  Ti\cite{Muscat1999}.}

Phonon frequencies are presented in
table~\ref{table:prim-frequencies}.  This shows clearly that
\textit{(a)} the rutile structure is predicted to suffer from a phonon
instability within PBE and \textit{(b)} that freezing the semicore
states has the effect of eliminating the instability and yields all
real frequencies. As the atomic reference configuration
is the same as used by Gr\"{u}nebohm \textit{et al.} it is highly
likely that this explains the failure to reproduce the instability in
their calculation.  The change in frequencies when the UF3
pseudopotentials are used at the lattice parameters of the UF2
calculation is much smaller than the UF2-UF3 change. This proves 
that the discrepancy between Gr\"{u}nebohm's
result and ours is a direct consequence of the omission of electronic
polarizability and hybridisation of the \ce{Ti} $3s$ and $3p$ states and not
an indirect consequence of the volume change.

The accuracy of the pseudopotentials is confirmed by the LDA results
which are in excellent agreement with previous work including FP-LAPW
calculations. The similarity of the UF2 results to the nearly
all-electron UF1 values for both LDA and PBE demonstrates that the UF2
configuration contains all of the electronic freedom required for
accurate lattice dynamics. This confirms the correctness of the
``small-core'' pseudopotential results of
Ref.~\onlinecite{Montanari2002} where it was demonstrated that
all-electron LCAO methods gave the same A$_{2u}$ instability as the
plane-wave pseudopotential calculations.  We note one discrepancy with
the prior results of Ref.~\onlinecite{Montanari2002}, namely that the
B$_{2u}$ and E$_{u}$ modes also exhibit an instability in addition to
the A$_{2u}$ instability previously described.

\section{Mode softening at q=(1/2,1/2,1/4)}

\begin{table}

\caption{\label{table:hhq-frequencies}  Frequencies of the lowest mode at q=(0.5,0.5,0.25)
calculated using using a $\sqrt{2} \times \sqrt{2} \times 4$ supercell (cm$^{-1}$).}

\begin{ruledtabular}
\begin{tabular}{lddd}
     &      \multicolumn{1}{c}{UF1}  &  \multicolumn{1}{c}{UF2}  &       \multicolumn{1}{c}{UF3} \\ \hline
PBE  &    77.1i  & 72.6i &      9.8i \\
LDA  &    50.0   & 55.1  &      74.8
\end{tabular}
\end{ruledtabular}
\end{table}

Turning to the soft acoustic mode frequencies, we performed supercell
phonon calculations using a $\sqrt{2} \times \sqrt{2} \times 4$
supercell and diagonalized at the commensurate wavevector
q=(0.5,0.5,0.25). Frequencies of the lowest acoustic modes are
presented in table~\ref{table:hhq-frequencies}.  Clearly the UF3
pseudopotential makes a substantial error in this
mode, which is stiffened when the effect of semicore polarization is
excluded.  As with the $\Gamma$ -point phonons, there is little
difference between the unfrozen core UF2 compared with the nearly
all-electron UF1 frequencies.  The UF1 and UF2 LDA results are in
good agreement with our earlier calculations in Ref.~\onlinecite{mitev2010}.
However our current calculations predict that within the PBE
approximation this acoustic mode has imaginary frequency for all three
pseudopotential configurations considered.  As with the Gamma point
results, we would expect the UF3 case to most closely reproduce the
results of Ref.~\onlinecite{Grunebohm2011}.

Gr\"{u}nebohm \textit{et al.} state that ``we could not reproduce the
destabilization of the system along an acoustic phonon mode under
(110) strain found in Ref. 8\footnotemark[3], for which an energy gain of several meV
has been predicted, even if we use a $2 . \sqrt{2} \times 2 . \sqrt{2}
\times 4$ supercell, which is commensurable with the displacement
pattern of this mode''. 
\footnotetext[3]{Our Ref. 6.}

However the geometry optimization method as employed by Gr\"{u}nebohm \textit{et al.}
is unsuitable for finding the soft displacement pattern. 
Our experience is that quasi-Newton type optimization methods usually fail any attempt to identify 
a shallow soft mode, or at best converge extremely slowly. Discovery of a
single set of atomic displacements which slightly lowers the energy in a high-dimensional 
search space of otherwise stiff directions is a highly computationally challenging procedure. Several factors contribute to the difficulty. 
First is the high dimensionality of the search space, 576 for the $\sqrt{2} \times \sqrt{2} \times 4$ cell employed (which is 
actually 4 times larger than strictly necessary to represent this soft eigenvector). Within this space
the desired eigenvector is represented by a single direction.

The second factor is the shape and depth of the anharmonic potential, which has a well depth of only 2.5 meV 
(see Fig. 5 of Ref.~\onlinecite{mitev2010})  and a tiny gradient along the mode. For small displacement this yields a typical force on each \ce{Ti} of 0.01 eV/A, with a maximum of 0.03 eV/A. 
Consequently the configuration at the initial points along the optimization would be deemed 
already converged according to the tolerance of 0.01 eV/A used by Gr\"{u}nebohm \textit{et al}. In our experience
a tolerance of 0.001 eV/A or even smaller is essential in such circumstances.

The third factor is the difficulty of finding a suitable starting configuration for the optimization.
In a supercell of perfect rutile the gradient in every direction is zero, so an initial perturbation is required to
break the supercell symmetry. A randomized displacement pattern must also excite the stiff 
modes and yield a much larger increase in the energy than the decrease being sought. The resulting geometry
optimization almost always fails to find the soft mode displacement.
We tested this for the soft acoustic mode in rutile \ce{TiO2}, using the UF2 pseudopotential, a force tolerance of 0.001 eV/A
and as expected the optimization returned to the symmetric supercell configuration. 

From the UF3 phonon frequency results in table~\ref{table:hhq-frequencies} we would expect an even shallower 
potential and even tinier gradient as a result of the additional stiffening resulting from the neglect of semi-core polarization.  
It is therefore doubly unsurprising that Gr\"{u}nebohm \textit{et al.}'s attempt to search for the soft displacement using this method was unsuccessful. 
This is absence of evidence for the soft mode, not evidence of absence.

In our experience the only reliable way to explore a soft mode using geometry optimization is to first perform a
phonon calculation, and to identify the soft eigenvector by diagonalization of a dynamical matrix.  
This eigenvector can be used as an initial perturbation from the symmetric supercell. The geometry optimization must
use very tight force and energy tolerances, but is usually able to proceed without interference
from the stiff modes.

\section{Conclusion}
We have demonstrated that the finding of two of us that
rutile \ce{TiO2} is mechanically unstable to a ferroelectric A$_{2u}$
soft mode in the PBE-GGA approximation is robust.
Gr\"{u}nebohm \textit{et al.}'s contrary finding can be attributed to an error
in the PAW pseudopotentials they used, namely the frozen-core
approximation applied to the \ce{Ti} $3s$ and $3p$ electrons. Our convergent
series of pseudopotentials to a highly accurate ``nearly all
electron'' limit demonstrates that the effect of freezing \ce{Ti} $3s$ and $3p$
is to stabilise the PBE rutile structure of \ce{TiO2}.  In fact this
error almost cancels the PBE instability to yield $\Gamma$ point
frequencies very similar to the LDA result but the cancellation is
less complete elsewhere in the Brilloun Zone.

We further argue that the geometry optimisation methods with convergence tolerances
reported by Gr\"{u}nebohm \textit{et al.} are not capable of identifying
shallow soft modes in the Brillouin zone. Our result\cite{mitev2010}
of a soft mode at q=(1/2,1/2,1/4) is also robustly confirmed.

Thus the situation is that in the incipient ferroelectric rutile-\ce{TiO2}
the LDA and PBE-GGA approximations to density functional theory yield
qualitatively distinct descriptions. Within PBE-GGA, rutile \ce{TiO2} is
predicted to be ferroelectric with an equilibrium geometry that is
unstable with respect to displacement along a number of phonon
eigenvectors\footnotemark[2]. Within the LDA, rutile-\ce{TiO2} is stable and a reasonable
description of its dielectric response and lattice dynamics is
obtained.

\begin{acknowledgments}
  We thank the EPSRC for support, of the UK national
  supercomputing facility (HECToR) under grant EP/F036809/1. Other
  calculations used STFC's e-Science facility. K.R. would also like to
  thank Professor Chris J. Pickard for introducing him to the concept of
  all-electron pseudopotentials.
\end{acknowledgments}

\end{document}